\documentclass[epj,nopacs]{svjour}
\usepackage{graphics}
\usepackage{multirow}
\usepackage{amstext}
\usepackage{amsmath}
\begin{document}
\title{Diffraction and Total Cross-Section\\ at the Tevatron and the LHC}
\author{M. Deile\inst{1} \and G. Anelli\inst{1} \and A. Aurola\inst{2}
\and V. Avati\inst{1} \and  V. Berardi\inst{3} \and  U.Bottigli\inst{4} 
\and M. Bozzo\inst{5}
\and E. Br\"{u}cken\inst{2} \and A. Buzzo\inst{5} \and M.~Calicchio\inst{3} 
\and F. Capurro\inst{5}
\and M.G. Catanesi\inst{3} \and M.A.Ciocci\inst{4} \and S. Cuneo\inst{5} 
\and C. Da Vi\`{a}\inst{6} \and E. Dimovasili\inst{1} \and K. Eggert\inst{1} 
\and M. Er\"{a}luoto\inst{2} \and F. Ferro\inst{5}
\and A. Giachero\inst{5} \and J.P. Guillaud\inst{7} \and J. Hasi\inst{6} 
\and F. Haug\inst{1} \and J. Heino\inst{2} 
\and T. Hilden\inst{2} \and P. Jarron\inst{1} \and J.~Kalliopuska\inst{2} 
\and J. Ka\v{s}par\inst{8} \and J. Kempa\inst{9}
\and C. Kenney\inst{10} \and A. Kok\inst{6} \and V. Kundr\'at\inst{8} 
\and K. Kurvinen\inst{2} 
\and S. Lami\inst{4} \and J. L\"{a}ms\"{a}\inst{2}
\and G. Latino\inst{4} \and R. Lauhakangas\inst{2} \and J. Lippmaa\inst{2} 
\and M. Lokaj\'{\i}\v{c}ek\inst{8} 
\and M. LoVetere\inst{5} \and D. Macina\inst{1} \and M. Macr\'{\i}\inst{5} 
\and M. Meucci\inst{4}
\and S.~Minutoli\inst{5} \and A. Morelli\inst{5} \and P. Musico\inst{5} 
\and M. Negri\inst{5}
\and H. Niewiadomski\inst{1} \and E. Noschis\inst{1} \and J. Ojala\inst{2} 
\and F. Oljemark\inst{2}
\and R. Orava\inst{2} \and M. Oriunno\inst{1} \and K. \"{O}sterberg\inst{2} 
\and R.Paoletti\inst{4} \and S. Parker\inst{11} \and A.-L. Perrot\inst{1}
\and E. Radermacher\inst{1} \and E. Radicioni\inst{3}
\and E. Robutti\inst{5} \and L. Ropelewski\inst{1} \and G. Ruggiero\inst{1}
\and H. Saarikko\inst{2} \and G.Sanguinetti\inst{4} \and A. Santroni\inst{5} 
\and S. Saramad\inst{1} 
\and F. Sauli\inst{1} \and A.Scribano\inst{4} \and G. Sette\inst{5}
\and J. Smotlacha\inst{8} 
\and W. Snoeys\inst{1} \and C. Taylor\inst{12} \and A. Toppinen\inst{2} 
\and N.Turini\inst{4}
\and N. Van Remortel\inst{2} \and L. Verardo\inst{5} \and A.~Verdier\inst{1} 
\and S. Watts\inst{6} \and J. Whitmore\inst{13} 
}
\institute{CERN, Gen\`{e}ve, Switzerland 
\and Helsinki Institute of Physics and University of Helsinki, Finland
\and INFN Sezione di Bari and Politecnico di Bari, Bari, Italy
\and  Universit\`{a} di Siena and Sezione INFN-Pisa, Italy
\and Universit\`{a} di Genova and Sezione INFN, Genova, Italy
\and Brunel University, Uxbridge, UK
\and LAPP Annecy, France
\and Academy of Sciences of the Czech Republic and Institute of Physics, 
Praha, Czech Republic
\and Warsaw University of Technology, Plock, Poland
\and Molecular Biology Consortium, SLAC, USA
\and University of Hawaii, USA
\and Case Western Reserve University, Dept. of Physics, Cleveland, OH, USA
\and Penn State University, Dept. of Physics, University Park, PA, USA}
\date{Received: date / Revised version: date}
\abstract{
At the Tevatron, the total $p\bar{p}$ cross-section has been measured by
CDF at 546\,GeV and 1.8\,TeV, and by E710/E811 at 1.8\,TeV. The two results
at 1.8\,TeV disagree by 2.6 standard deviations, introducing big uncertainties
into extrapolations to higher energies. 
At the LHC, the TOTEM collaboration is preparing to resolve the ambiguity by
measuring the total $pp$ cross-section with a precision of about 1\,\%.
Like at the Tevatron experiments, the luminosity-independent method based on
the Optical Theorem will be used. 
The Tevatron experiments have also performed a vast range of studies about
soft and hard diffractive events, partly with antiproton tagging by Roman Pots,
partly with rapidity gap tagging. At the LHC, the combined CMS/TOTEM 
experiments will carry out their diffractive programme with 
an unprecedented rapidity coverage and Roman Pot 
spectrometers on both sides of the interaction point. 
The physics menu comprises detailed studies of soft diffractive differential 
cross-sections, diffractive structure functions, rapidity gap survival and
exclusive central production by Double Pomeron Exchange.
} 
\maketitle
\section{Introduction}
\label{sec_intro}
\begin{figure*}
\mbox{
\resizebox{0.6\textwidth}{!}{
  \includegraphics{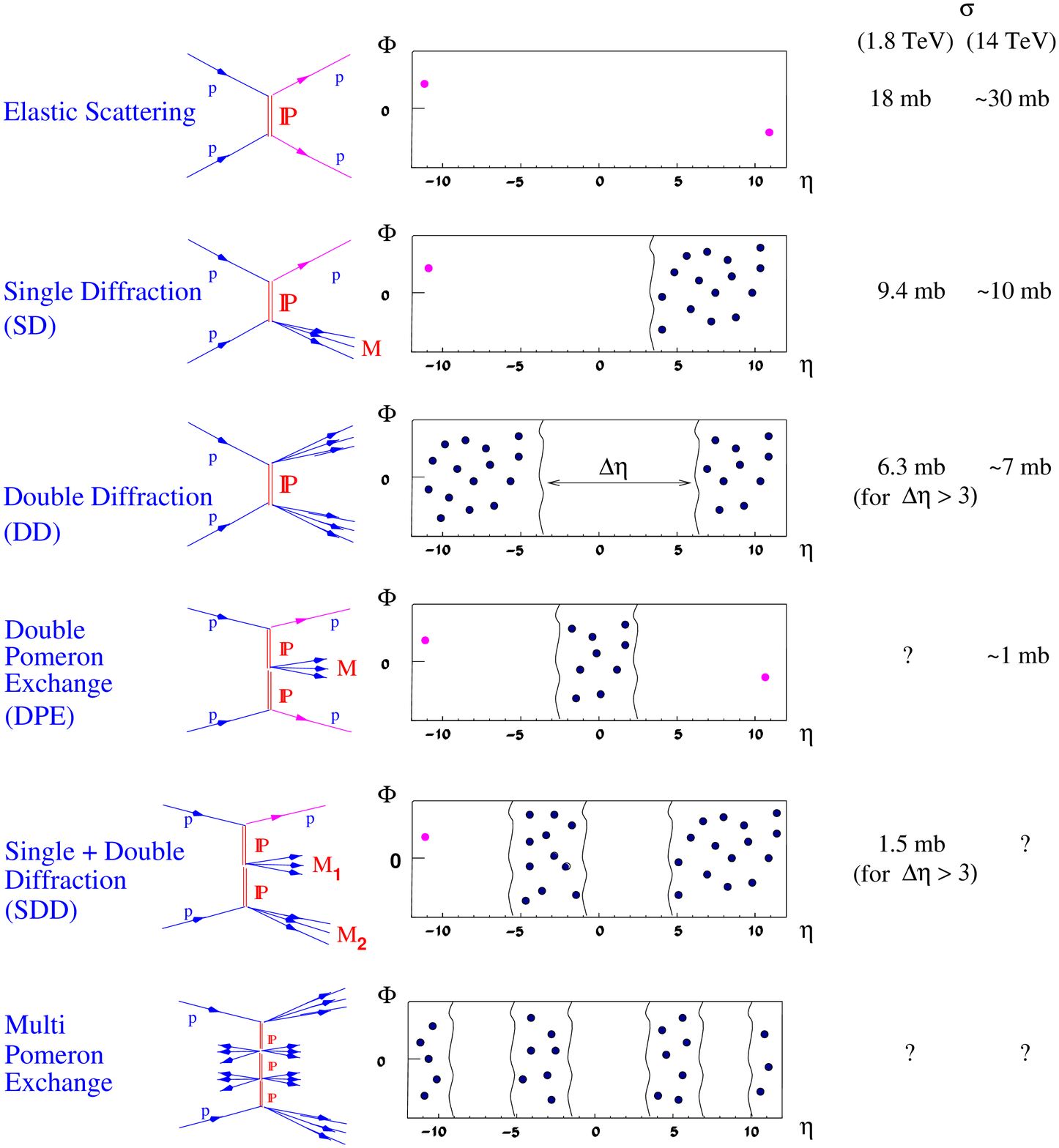}
}
\hspace*{-3mm}
\raisebox{5cm}{
\scalebox{0.95}{
\begin{tabular}{|c|cccc|}
\multicolumn{5}{l}{TOTEM(+CMS) Diffractive Running Scenarios}\\
\multicolumn{5}{l}{($k$: number of bunches)}\\
\multicolumn{5}{l}{($N$: number of protons per bunch)}\\
\hline
Scen. & $\beta^{*}$[m] & $k$ & $N/10^{11}$   & $\mathcal{L} [{\rm cm^{-2}s^{-1}}]$ \\
\hline
1        & 1540        & 43   & 0.3          & $1.6\times10^{28}$\\
2        & 1540        & 156  & 0.6$\div$1.15& $2.4\times10^{29}$\\
3        &   18        & 2808 & 1.15         & $3.6\times10^{32}$\\
4        &   90        & 936  & 1.15         & $2\times10^{31}$\\
5        &   0.5       & 2808 & 0.3          & $10^{33}$\\
\hline
1        & \multicolumn{4}{l|}{low $|t|$ elastic, $\sigma_{T}$, min. bias, soft diffract.}\\
2        & \multicolumn{4}{l|}{diffraction}\\
3        & \multicolumn{4}{l|}{large $|t|$ elastic}\\
4        & \multicolumn{4}{l|}{hard diffraction, large $|t|$ elastic}\\
         & \multicolumn{4}{l|}{({\it under study}~\cite{karstenblois05})}\\
5        & \multicolumn{4}{l|}{rare diffractive processes, \it for later}\\
\hline
\end{tabular}
}}}
\caption{Left: diffractive process classes and their cross-sections at Tevatron and LHC.
Right: running scenarios for diffractive physics at LHC; for more details 
see~\cite{viennaproc}.}
\label{fig_eventclasses}   
\end{figure*}

Elastic and diffractive scattering (see Fig.~\ref{fig_eventclasses}, left) 
represent a significant 
fraction (44\,\% at both $\sqrt{s}=1.8$\,TeV and 14\,TeV) 
of the total $pp$ or $p\bar{p}$ cross-section. 
Many details of these processes with close ties to proton
structure and low-energy QCD are still not understood. 
The main signature -- large gaps in the scattering products' rapidity 
distribution due to exchange of colour singlets between the interacting 
protons -- leads to the requirement of a good rapidity coverage up to the 
very forward region. This is also needed for the detection of high-$p_{T}$
particles and jets from hard diffractive events 
-- i.e. those with hard partonic subprocesses -- which convey 
information about the partonic structure of the colour singlet 
(a.k.a. ``Pomeron'') exchanged.
A big fraction of diffractive events exhibits
surviving (``leading'') protons at very small scattering angles which
can be detected in Roman Pot detectors far away from the interaction point.

Another purpose of high-coverage detector systems is the 
luminosity-independent determination of the total cross-section based 
on the Optical Theorem which requires the measurement of the 
total elastic and inelastic rates and the extrapolation of the nuclear elastic
scattering cross-section $d\sigma/dt$ to zero momentum transfer, $t = 0$, 
as explained in Section~\ref{sec_sigmatot}.

The Tevatron experiments CDF~\cite{cdfelastic}, E710~\cite{e710} 
and its very similar successor E811~\cite{e811} had Roman Pots
on both sides of the interaction points for detecting elastically scattered 
protons. For diffractive physics, only the antiproton side had 
enough dispersion for measuring leading particle momenta with
Roman Pot spectrometers.
The rapidity coverage for measuring the inelastic rate ranged from 
5.2 to 6.5 at E710/811 and from 3.2 to 6.7 at CDF. For tagging diffractive
events by their rapidity gaps, additional central detectors were available 
extending the coverage to $\pm$(3.8$\div$6.5) for E710 and 0$\div(\pm$5.9) 
(7.5) for CDF in Run I (Run II). 

At D\O, a double-arm Roman Pot spectrometer (FPD) was installed for 
Run II~\cite{d0forward},
allowing to measure elastic and diffractive processes with (anti-) proton
acceptance on both sides of the interaction point. In Run I, rapidity gap
tagging was possible for $|\eta| < 5.9$.

The TOTEM experiment~\cite{totemtdr} at the LHC will have Roman Pot stations 
at 147\,m and
at 220\,m from the interaction point, on both sides. The inelastic event rate
will be measured in a rapidity interval from 3.1 to 6.5. For diffractive 
physics, TOTEM will collaborate with CMS, resulting in a
rapidity coverage from 0 to $\pm$6.5. 


\section{Elastic $\mathbf{pp}$ and $\mathbf{p\bar{p}}$ Scattering}

\begin{figure*}[ht]
\begin{center}
\resizebox{0.9\textwidth}{!}{
  \includegraphics{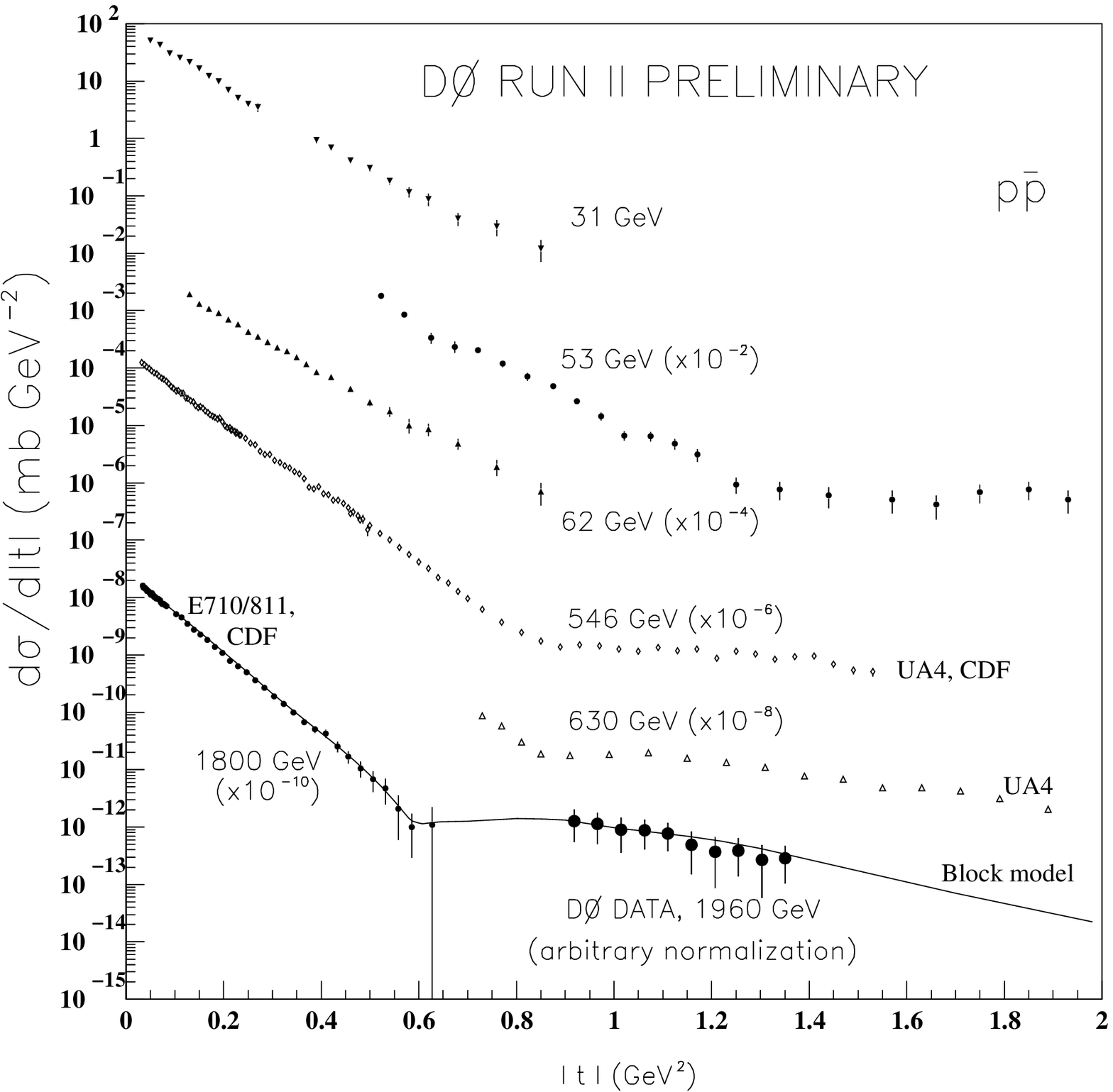} 
  \hspace*{10mm}
  \includegraphics{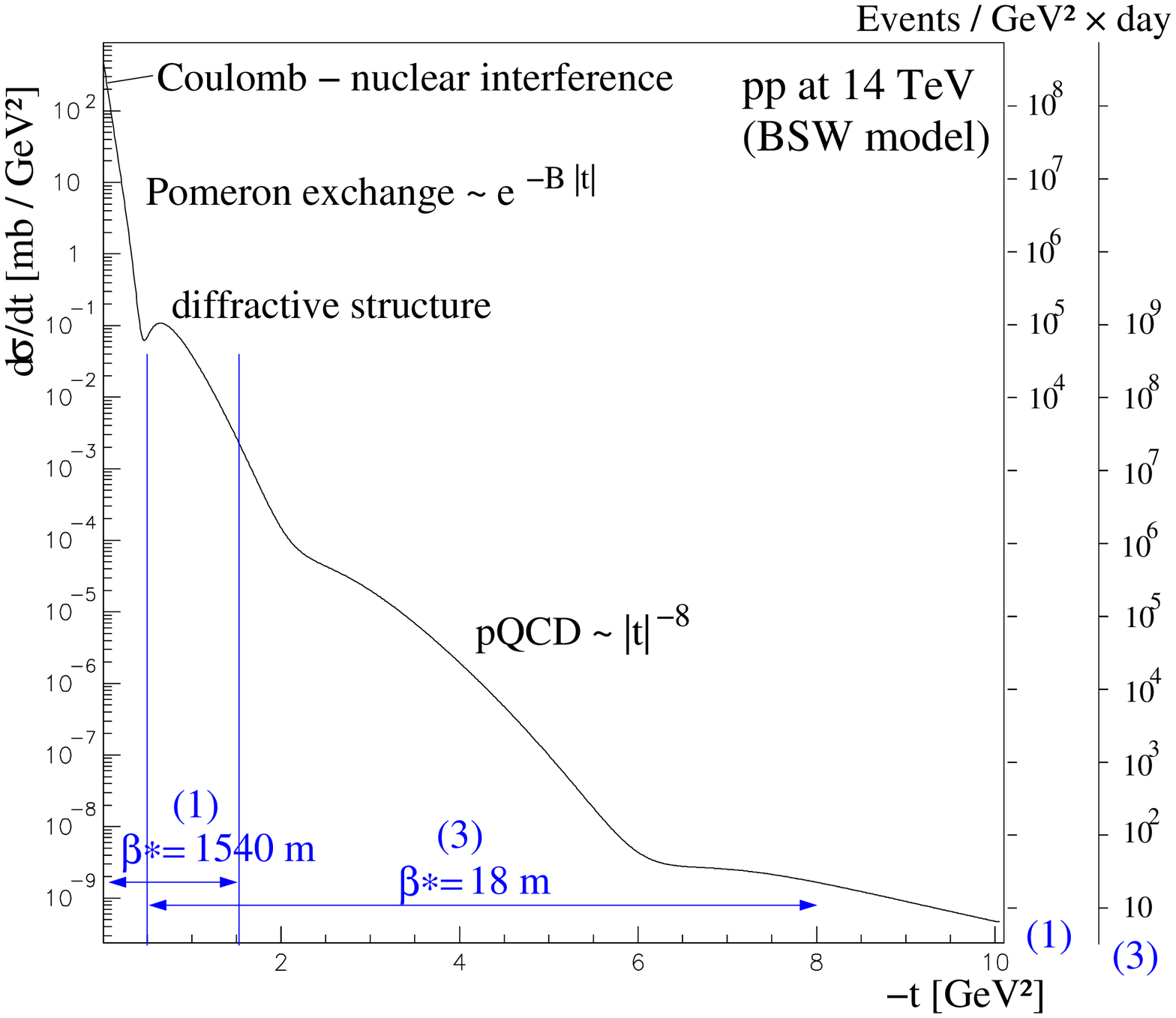} 
}
\end{center}
\caption{Left: elastic $p\bar{p}$ scattering from ISR to Tevatron (taken from 
\cite{d0proc}); right: prediction for elastic $pp$ scattering at LHC; the 
one-day statistics on the right-hand scales correspond to the running 
scenarios 1 and 3 (defined in Fig.~\ref{fig_eventclasses}).}
\label{fig_elastic}
\end{figure*}
The elastic scattering cross-section $d\sigma/dt$ is characterised by 
several $t$-regions with different behaviour (see Fig.~\ref{fig_elastic}):
\begin{itemize}
\item The Coulomb region where elastic scattering is dominated by photon 
exchange; this region lies at 
$|t| < 1.2 \times 10^{-3}\,{\rm GeV}^{2}$ for $\sqrt{s}$=546\,GeV, 
$|t| < 0.9 \times 10^{-3}\,{\rm GeV}^{2}$ for $\sqrt{s}$=1.8\,TeV, 
and  $|t| < 6.5\times 10^{-4}\,{\rm GeV}^{2}$ for $\sqrt{s}$=14\,TeV.
\item The nuclear/Coulomb interference region, where the cross-section is
given by
\begin{multline}
\label{eqn_elasticinterference}
\frac{d\sigma_{el}}{dt} = \pi |f_{C} {\rm e}^{-i \alpha \phi(t)} 
+ f_{N}|^{2} \\
= \pi \left|-\frac{2 \alpha G^{2}(t)}{|t|} 
{\rm e}^{-i \alpha \phi(t)}
+ \frac{\sigma_{tot}}{4 \pi} 
|i + \rho| {\rm e}^{-B |t| / 2}\right|^{2} \:.
\end{multline}
Here, $G(t)$ is the electromagnetic form factor of the proton, $\rho$ 
the ratio between real and imaginary part of the forward nuclear elastic
amplitude,
\begin{equation}
\rho = \frac{\mathcal{R}[f_{el}(0)]}{\mathcal{I}[f_{el}(0)]} \:,
\end{equation}
and $\phi$ is the relative phase between the nuclear and Coulomb 
amplitudes. E710 and E811~\cite{e710,e811} have measured $\rho$ and $B$ in 
this region
(see Table~\ref{tab_elastic_tevatron}), using the West-Yennie parameterisation
for $\phi(t)$~\cite{westyennie}.
The interest of $\rho$ lies in its predictive power for $\sigma_{tot}$ at 
higher energies via the dispersion relation 
\begin{equation}
\rho(s) = \frac{\pi}{2 \sigma_{tot}(s)} \frac{d\sigma_{tot}}{d\ln s}
\end{equation}
\item The ``single-Pomeron exchange'' region with a cross-section 
$d\sigma / dt \propto {\rm e}^{-B\,|t|}$. The parameter $B$ was measured by
several Tevatron experiments (Table~\ref{tab_elastic_tevatron}).
\item A region with diffractive minima which move to lower $|t|$ as the 
energy increases (Fig.~\ref{fig_elastic}, left).
\item The triple-gluon exchange region at high $|t|$ described by 
perturbative QCD and showing a cross-section proportional to $|t|^{-8}$.
\end{itemize}

\begin{table}[h!]
\caption{Elastic scattering at the 
Tevatron~\cite{e710,e811,cdfelastic,d0proc}}
\label{tab_elastic_tevatron} 
\begin{tabular}{|llll|}
\hline
$\sqrt{s}$ & Exp. & $t$-range [GeV$^{2}$] & $B [{\rm GeV}^{-2}]$, $\rho$ \\
\hline
546\,GeV & CDF & 0.025 $\div$ 0.08 & $B = 15.28 \pm 0.58$\\
1.8\,TeV & CDF & 0.04 $\div$ 0.29  & $B = 16.98 \pm 0.25$\\
         & E710& 0.034 $\div$ 0.65 & $B = 16.3 \pm 0.3$\\
         &     & 0.001 $\div$ 0.14 & $B = 16.99 \pm 0.25$\\
         &     &                   & $\rho = 0.140 \pm 0.069$\\
         & E811& 0.002 $\div$ 0.035&using $\langle B \rangle_{\rm CDF, E710}$\\
         &     &                   & $\rho = 0.132 \pm 0.056$\\
1.96\,TeV& D\O  & 0.9 $\div$ 1.35   &          --             \\
\hline
\end{tabular}
\end{table}
The TOTEM experiment at LHC will cover the $|t|$-range from 
$2 \times 10^{-3}\,$GeV$^{2}$ to 8\,GeV$^{2}$ (Fig.~\ref{fig_elastic}, right) 
with two running scenarios with special beam optics and different 
luminosities (scenarios 1 and 3 (or 4) in Fig.~\ref{fig_eventclasses}, right).
For details of
the $t$-acceptances of the scenarios see Ref.~\cite{viennaproc}.
The minimum $|t|$-value corresponds to a distance of 1.3\,mm = 
10 $\sigma_{beam}$ + 0.5\,mm between the 
Roman Pot at 220\,m and the beam centre.
Reaching the Coulomb-nuclear interference region to measure $\rho$ will be
attempted either by approaching the beam closer with the Roman Pot or by
operating the LHC at $\sqrt{s} \le 6\,$TeV (see Fig.~4 in ~\cite{viennaproc}).


\section{Total $\mathbf{pp}$ and $\mathbf{p\bar{p}}$ Cross-Section}
\label{sec_sigmatot}
The total $\mathbf{pp}$ or $\mathbf{p\bar{p}}$ cross-section is related to 
nuclear elastic forward scattering
via the Optical Theorem which can be expressed as
\begin{equation}
\label{eqn_optical}
\mathcal{L} \sigma_{tot}^{2} = \frac{16 \pi}{1 + \rho^{2}} \cdot
\left.\frac{dN_{el}}{dt} \right|_{t=0} \:.
\end{equation}
With the additional relation
\begin{equation}
\label{eqn_totalrate}
\mathcal{L} \sigma_{tot} = N_{el} + N_{inel}
\end{equation}
one obtains a system of 2 equations which can be resolved for $\sigma_{tot}$ 
or $\mathcal{L}$ independently of each other:
\begin{eqnarray}
\sigma_{tot} &=& \frac{16 \pi}{1 + \rho^{2}} \cdot
\frac{dN_{el}/dt |_{t=0}}{N_{el} + N_{inel}}\:,\label{eqn_sigmatot}\\
\mathcal{L} &=& \frac{1 + \rho^{2}}{16 \pi} \cdot 
\frac{(N_{el} + N_{inel})^{2}}{dN_{el}/dt |_{t=0}} 
\end{eqnarray}
Hence the quantities to be measured are:
\begin{itemize}
\item the nuclear part of the elastic cross-section extrapolated to $t = 0$;
\item the total elastic and inelastic rate, the latter consisting of 
diffractive (18\,mb at LHC) and minimum bias (65\,mb at LHC) events.
\end{itemize}
The $\rho$ parameter has to be taken from external knowledge unless it can 
be measured from elastic scattering in the interference region between 
nuclear and Coulomb scattering. 
CDF have measured $\sigma_{tot}$ at 546\,GeV and 1.8\,TeV using 
Eqn.~(\ref{eqn_sigmatot}) with $\rho=0.15$~\cite{cdfelastic} 
(see Table~\ref{tab_sigmatot}). Their measurement at 546\,GeV agrees with the
value from UA4~\cite{ua4}.
E710 and E811 have determined $\rho$ and 
$\sigma_{tot}$ simultaneously at 1.8\,TeV~\cite{e710,e811} by 
combining Eqns.~(\ref{eqn_optical}) and (\ref{eqn_totalrate}) 
with~(\ref{eqn_elasticinterference}). Their result for $\sigma_{tot}$ differs
from CDF's number by 2.6 standard deviations. The origin of the discrepancy is
unknown. 

\begin{table}[h!]
\vspace*{-3mm}
\caption{Measurements of the total $pp$ or $p\bar{p}$ cross-section for 
$\sqrt{s} \ge 546\,$GeV and expectations for the LHC.}
\label{tab_sigmatot} 
\begin{tabular}{|lll|}
\hline
$\sqrt{s}$ & Experiment & $\sigma_{tot}$ [mb] \\
\hline
546\,GeV & UA4 & $61.9~ \pm 1.5~$\\
         & CDF & $61.26 \pm 0.93$\\
1.8\,TeV & CDF & $80.03 \pm 2.24$\\
         & E710& $72.8~ \pm 3.1~$\\
         & E811& $71.42 \pm 2.41$\\
14\,TeV  & (extrapolation~\cite{compete} to LHC) & $111.5 \pm 1.2^{+4.1}_{-2.1}$\\
         & TOTEM & $~~~?~~ \pm 1~~$ \\
\hline
\end{tabular}
\end{table}
\begin{figure}
\begin{center}
\resizebox{0.45\textwidth}{!}{
  \includegraphics{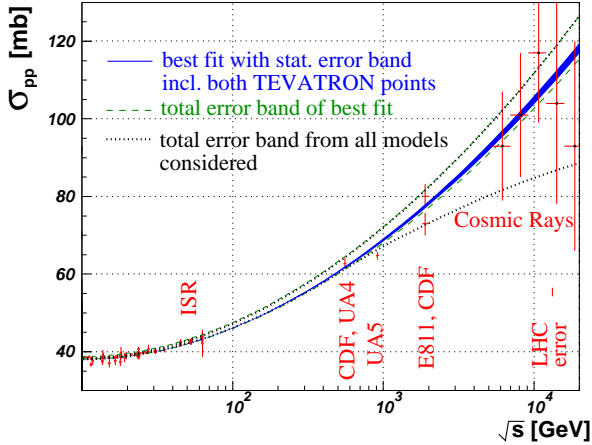}
}
\end{center}
\caption{COMPETE fits~\cite{compete} to all available $pp$ and $p\bar{p}$ 
scattering data with statistical (blue solid) and total (green dashed) error bands, the
latter taking into account the Tevatron ambiguity. The outermost curves (dotted) give 
the total error band from all parameterisations considered.}
\label{fig_sigmatot}   
\end{figure}
TOTEM will follow the same method as CDF. The total expected uncertainty 
of 1\,\% after 1 day of taking data at 
$\mathcal{L} = 1.6\times 10^{28}$\,cm$^{-2}$\,s$^{-1}$ will have the 
following contributions (combined in quadrature):
\begin{itemize}
\item The statistical errors of $N_{el} + N_{inel}$ and $dN_{el}/dt |_{t=0}$
are negligible: 0.01\,\% and 0.07\,\% respectively.
\item The systematic error of the total rate stems primarily from trigger 
losses and amounts to 0.8\,\%.
\item The systematic error of the extrapolation of the elastic cross-section
to $t = 0$ is dominated by the theoretical uncertainty of the functional form
(0.5\,\%). The next-to-leading contributions come from beam energy,
alignment and crossing-angle uncertainties (each typically 0.1\,\%).
\item If $\rho$ cannot be measured, the uncertainty in its prediction
(e.g. $\rho = 0.1361\pm0.0015^{+0.0058}_{-0.0025}$~\cite{compete}) will 
contribute another 0.2\,\%. 
\end{itemize}
The ATLAS collaboration proposes~\cite{atlasloi} to extract the four 
parameters $\sigma_{tot}$, 
$\rho$, $B$ and $\mathcal{L}$ from a fit to (\ref{eqn_elasticinterference}) 
and using $dN/dt = \mathcal{L} d\sigma/dt$. The main difficulties of this 
approach lie in reaching
low enough t-values ($-t < 6 \times 10^{-4}$\,GeV$^{2}$) and in the uncertainty
of the phase $\phi$.

\section{Diffraction}
At Tevatron, a vast number of studies on soft and hard diffraction has been
carried out (see Table~\ref{tab_diffractiontevatron} for a brief overview).

\begin{table}[h!]
\caption{The diffractive programmes of the Tevatron experiments, the methods
for tagging diffractive events, and the coverage in kinematic variables ($t$ is
given in units of GeV$^{2}$) The abreviations for the diffractive event classes
are defined in Fig.~\ref{fig_eventclasses} (left).}
\label{tab_diffractiontevatron} 
\begin{tabular}{|llll|}
\hline
Exp., Run & Tagging        & Coverage & Physics \\
\hline
E710       & rap. gap             & $3.8 < |\eta| < 6.5$ & 
\multirow{3}{2cm}{\hspace{-3mm}$\biggr\}$soft SD}  \\
\small\cite{e710} & leading $\bar{\rm p}$& $0.05 < -t < 0.11$   &          \\
           &                      & $\xi < 0.01$         &          \\
\hline
CDF I,0    & rap. gap             & $|\eta| < 6.7$       & 
\multirow{3}{2cm}{\hspace{-3mm}$\biggr\}$soft SD}  \\
\small\cite{cdfelastic} & leading $\bar{\rm p}$& $-t < 0.4$           &         \\
           &                      & $\xi < 0.2$          &          \\
\hline
CDF IA,B   & rap. gap             & $|\eta| < 5.9$       & 
\multirow{5}{2cm}{\hspace{-3mm}$\Biggr\}$\parbox{20mm}{soft SD, DD, DPE, SDD\\
hard diffract.:\\ dijets, W, b$\bar{\rm b}$, J/$\Psi$}}  \\
\small\cite{cdfIAB} & no RP                &                      &         \\
\cline{1-3}
CDF IC     & rap. gap             & $|\eta| < 5.9$       & \\
\small\cite{cdfIC}  & leading $\bar{\rm p}$& $-t < 1$             &         \\
           &                      & $0.03 < \xi < 0.1$   &    \\
\hline  
CDF II     & rap. gap             & $|\eta| < 7.5$       & 
\multirow{3}{2cm}{\hspace{-3mm}$\biggr\}$\parbox{20mm}{diffr. struct. funct.,
search for excl. DPE}}\\
\small\cite{cdfII} & leading $\bar{\rm p}$& $-t < 2$             &         \\
           &                      & $0.02 < \xi < 0.1$   &    \\
\hline
D\O~I      & rap. gap             & $|\eta| < 5.9$       & 
\multirow{2}{2cm}{\hspace{-3mm}$\biggr\}$\parbox{20mm}{hard diffr.:\\ 
dijets, W, Z}}     \\
\small\cite{d0I}   & no RP                &                      &         \\
\hline  
D\O~II     & rap. gap             & $|\eta| < 5.9$       & 
\multirow{3}{2cm}{\hspace{-3mm}$\biggr\}$\parbox{20mm}{all above with 
p, $\bar{\rm p}$ tagging}}     \\
\small\cite{d0proc} & lead. p, $\bar{\rm p}$ &  $0.8 < -t < 2$ &         \\
           &                      & any $\xi$           &          \\
\hline  
\end{tabular}
\end{table}
In Run I, diffractive events were tagged by their rapidity gaps and -- in 
some cases -- by a leading antiproton. Leading diffractive protons were
not detected. For the ongoing Run II on the other hand, D\O~has installed 
a double-arm proton- and antiproton spectrometer.

At TOTEM/CMS, for all diffractive processes (except DD) leading proton tagging 
is foreseen with the possibility of using rapidity gaps for redundancy.
With scenarios 1 and 2, used for soft and semi-hard diffraction, 
protons of all $\xi$ will be detected; the total
acceptance integrated over $t$ and $\xi$ is 95\,\%; the resolution in $\xi$ is about
$5\times10^{-3}$. Hard diffraction with its much smaller cross-sections
(e.g. 1\,$\mu$b for SD dijets at $\sqrt{s}=14\,$TeV) will be studied with scenario 4
where the total proton acceptance is about 65\,\%, and the $\xi$ resolution is about
$4\times10^{-4}$.

\subsection{Soft Diffraction}
At Tevatron, the total and differential soft diffractive cross-sections have 
been measured for 
the processes of SD (E710, CDF), DD and SDD (CDF), see 
Fig.~\ref{fig_eventclasses}. 
A central result of these cross-section studies is that the 
$t$ and $\xi$ dependences of the differential cross-sections conform to
the predictions of Regge Theory, but that the total normalisations measured are
suppressed, as also observed in hard diffraction (see below). 
With increasing $\sqrt{s}$ this suppression becomes more 
pronounced. The behaviour of the diffractive cross-sections at energies above 1.8\,TeV
is controversial between different models predicting it either to increase further or 
to remain constant~\cite{goulianos_sapeta}. From the ratios between $\sigma_{diff}$, 
$\sigma_{elast}$ and $\sigma_{tot}$, information about opacity and size of the proton
can be deduced.

In DPE, CDF's one-armed antiproton spectrometer
tagged the slightly wider ``inclusive'' event class $\rm \bar{p}p \rightarrow \bar{p}+X+Y$ 
where the proton is allowed to dissociate into a low-mass system Y with 
$m_{Y}^{2}\le8\,\rm GeV^{2}$. In the central diffractive system, masses 
up to a few $10^{2}$\,GeV were seen. At LHC, diffractive masses up to about 1.4\,TeV 
will be observable with sufficient statistics. Surviving protons will be detected on both
sides of the IP.

\subsection{Hard Diffraction}
\begin{figure}[h!]
\begin{center}
\resizebox{0.33\textwidth}{!}{
  \includegraphics{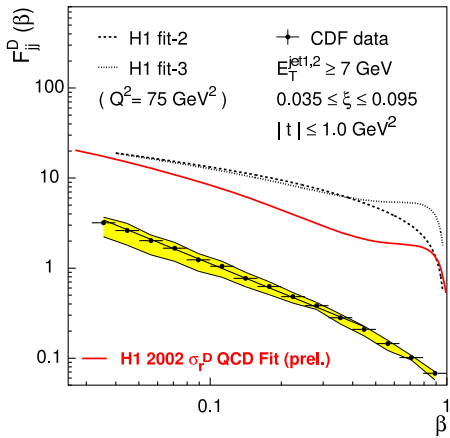}
}
\end{center}
\vspace*{-4mm}
\caption{Diffractive structure function for dijet production in
DDIS at H1 and in SD at CDF. The mean $(E_{T}^{jet})^{2}$ at 
CDF corresponds approximately to $Q^{2}$ at H1.}
\label{fig_heracdf}   
\end{figure}

A central result in diffraction at Tevatron is the breaking of QCD
factorisation, i.e. of the hypothesis that the cross-sections of 
hard diffractive processes can be written as a convolution
\begin{equation}
\label{eqn_qcdfact}
\sigma = \int d\beta\,dQ^{2}\,d\xi\,dt\,\,
\hat{\sigma}(\beta, Q^{2}, \xi, t) \,F_{2}^{D}(\beta, Q^{2}, \xi, t)
\end{equation}
of a parton-level cross-section $\hat{\sigma}$ and a process-independent 
diffractive structure function $F_{2}^{D}$. Comparing $F_{2}^{D}$ from 
dijet production in diffractive deep inelastic scattering (DDIS) at HERA 
with 
the result from single diffractive dijet production at Tevatron yields a
suppression of the latter by roughly a factor 10 (Fig.~\ref{fig_heracdf}).

This suppression of the diffractive cross-section
is independent of the hard subprocess, as can be seen by comparing for 
different partonic subprocesses the 
fractions of events showing rapidity gaps (Table~\ref{tab_gapfractions}).
They are all of the order 1\,\%. The variations are due to different sensitivities
to the gluon and quark components of the Pomeron and led to the determination of
the gluon fraction $f_{g}=0.59\pm0.14\pm0.06$ in agreement with HERA's 
$f_{g}=0.75\pm0.15$.

\begin{table}[h!]
\caption{Ratio R between the diffractive subsample (with rapidity gap) and 
all events for a given hard subprocess (j = jet, G = gap). $\sqrt{s}=1.8\,$TeV.}
\label{tab_gapfractions} 
\begin{tabular}{|llll|}
\hline
Process & Cuts & R [\%]& Exp.  \\
\hline
\multirow{4}{2.2cm}{SD: j + j + G} & $E_{T}>20$\,GeV, & 
      \multirow{2}{1.5cm}{$0.75\pm0.10$}& \multirow{2}{0.5cm}{CDF}\\
                    &  $\eta_{j} > 1.8$ &              &    \\
\cline{2-4}
                    & $E_{T}>12$\,GeV,                & 
      \multirow{2}{1.5cm}{$0.65\pm0.04$}& \multirow{2}{0.5cm}{D\O} \\
                    & $|\eta_{j}| > 1.6$&             &     \\
\hline
\multirow{4}{2.2cm}{DD: j + G + j} & $E_{T}>20$\,GeV, & 
      \multirow{2}{1.5cm}{$1.13\pm0.16$}& \multirow{2}{0.5cm}{CDF} \\
                    & $|\eta_{j}| > 1.8$&             &    \\
\cline{2-4}
                    & $E_{T}>30$\,GeV,                & 
      \multirow{2}{1.5cm}{$0.94\pm0.13$}& \multirow{2}{0.5cm}{D\O} \\
                    & $|\eta_{j}| > 1.6$, $\Delta\eta_{j}>4$&     &    \\
\hline
SD: W + G & 
      $E_{T}\hspace{-4mm}/~~, E_{Te} > 20$\,GeV& $1.15\pm0.55$& CDF\\
\cline{2-4}
~~$\rightarrow \rm e\,\nu$ + G  &   
      $E_{T}\hspace{-4mm}/~~, E_{Te} > 25$\,GeV &$0.89^{+0.20}_{-0.19}$ & D\O\\
\hline
\parbox{2.2cm}{SD: Z + G\\ ~~~$\rightarrow$ e\,e + G}& $E_{Te} > 25$\,GeV                  &$1.44^{+0.62}_{-0.54}$ & D\O\\
\hline
SD: b + G                 & $p_{Te}>9.5$\,GeV, & \multirow{2}{1.5cm}{$0.62\pm0.25$} & \multirow{2}{0.5cm}{CDF} \\
~~$\rightarrow$ e\,X + G  & $|\eta_{e}| < 1.1$ &               &     \\
\hline
SD: J/$\Psi$ + G          & $p_{T\mu}>2$\,GeV, & \multirow{2}{1.5cm}{$1.45\pm0.25$} & \multirow{2}{0.5cm}{CDF} \\
~$\rightarrow \mu^{+}\,\mu^{-}$ + G  & $|\eta_{\mu}| < 0.6$ &    &     \\
\hline
\end{tabular}
\vspace*{-3mm}
\end{table}
A possible explanation lies in the different initial states in DDIS and in
proton-antiproton diffraction. In the latter case, additional soft scattering
between the two initial hadrons can fill the rapidity gap and thus destroy
the signature used for identifying diffractive events. Hence the cross-section
in Eqn.~(\ref{eqn_qcdfact}) needs the ``gap survival probability'' $|S|^{2}$
as another convolution factor. $|S|^{2}$ was observed by CDF to decrease by a
factor 1.3$\div$2.4 from 630\,GeV to 1.8\,TeV and is expected to be further
reduced at LHC energies. The measurement of gap probabilities at the LHC will be 
an important input for the study of exclusive production processes discussed in the next
section.

At LHC, additional hard phenomena offering insight into proton structure are being 
explored, like exclusive SD into three jets,
pp $\rightarrow$ p + jjj, which would indicate a
{\it minimal Fock space} parton configuration $|qqq\rangle$ in the 
proton~\cite{colourtrans}. For a jet
threshold of 10\,GeV, a cross-section between 0.04 and 0.4\,nb is 
predicted, yielding 80 to 800 events per day at 
$\mathcal{L}=2\times 10^{31}\,{\rm cm^{-2}s^{-1}}$ (scenario 4).

\subsection{Exclusive Production by DPE}
A particularly interesting subclass of DPE events is exclusive central production,
characterised by only one single particle or a dijet in the diffractive system.
The vacuum quantum numbers of the two colliding colour singlets lead to selection
rules on spin $J$, parity $P$ and charge conjugation 
$C$~\cite{kmr_rules}:
\begin{equation}
\label{eqn_selectionrules}
J^{P} = 0^{+}, 2^{+}, 4^{+}; J_{z} = 0; C = +1
\end{equation}
(in the limit of $t = 0$). 
The $J_{z} = 0$ rule strongly suppresses gg$\rightarrow {\rm q}\bar{\rm q}$ 
background because of helicity conservation (this background would totally 
vanish for massless quarks). The rules can also be used for determining the 
quantum numbers of a new state observed. Table~\ref{tab_exclusiveDPE} lists some
examples for exclusive production. For exclusive dijet and $\chi_{c0}$ production,
CDF has seen event candidates and set upper limits on the cross-section. At LHC, these
processes should be well within reach using scenario 4. 
The observability of the $\chi_{b0}$ is 
doubtful because the branching ratio for its muonic decay is unknown 
(upper limit: $10^{-3}$).

\begin{table}[h!]  
\vspace*{-3mm}
\caption{Examples of exclusive DPE processes (p + p $\rightarrow$ p + X + p).
For cross-sections see e.g.~\protect\cite{kmr_petrov}. The numbers in square brackets
are experimental upper limits from CDF, Run II~\cite{cdfII}.}
\label{tab_exclusiveDPE}  
\scalebox{0.88}{  
\begin{tabular}{|c|c|c|c|}\hline 
Diffractive         & Decay channel & $\sigma(Tev.)\times$BR & $\sigma(LHC)\times$BR\\
system              &               &                        &           \\
\hline\hline
dijet               & \multirow{2}{5mm}{jj} &  0.97\,nb      & 7\,nb     \\
($E_{T}>10\,$GeV)   &                       & [$\le$1.1\,nb]      &           \\
\hline
$\chi_{c0}$    & $\gamma J/\psi \rightarrow \gamma \mu^{+} \mu^{-}$
                                    & 390\,pb                & 1.8\,nb \\
(3.4\,GeV)     &                                                   
                                    & [$\le$204\,pb]$^{1}$   &           \\
              & $\pi^{+} \pi^{-} K^{+} K^{-}$ 
                                    & 12\,nb                 & 54\,nb    \\
\hline
$\chi_{b0}$    & \multirow{2}{20mm}{$\gamma Y \rightarrow \gamma \mu^{+} \mu^{-}$}
                                    & \multirow{2}{12mm}{$\le$\,0.5\,pb}  & \multirow{2}{10mm}{$\le$4\,pb}   \\
(9.9\,GeV)     &                                              
                                    &                        &        \\
\hline
\multicolumn{4}{l}{\hspace*{-2mm}$^{1}$ scaled from CDF's rapidity range $\pm0.6$ to $\pm2.5$ used by 
KMRS~\cite{kmr_petrov}.\hspace*{-4mm}}
\end{tabular}
}
\vspace*{-3mm}
\end{table}
%
%
\begin{table}[h!]
\vspace*{-3mm}
\caption{Cross-sections for exclusive Higgs production in the SM and the MSSM
(examples)~\cite{KKMR}.
A mass resolution $\sigma(M) =$ 3\,GeV from the Roman Pot spectrometer is 
assumed.}
\label{tab_higgs} 
\scalebox{0.9}{
\begin{tabular}{|l|l|l|}
\hline
\textbf{SM,} $\mathbf{m_{H} = 120}$\,\textbf{GeV} & \multicolumn{2}{c|}{}  \\
\hline
$\sigma \times {\rm BR (H\rightarrow b\bar{b})}$ & 
\multicolumn{2}{c|}{~~2\,fb (S/B @ 30\,fb$^{-1}$ = 11/10)}\\
$\sigma \times {\rm BR (H\rightarrow WW^{*})}$ & 
\multicolumn{2}{c|}{0.4\,fb (S/B @ 30\,fb$^{-1}$ = ~8/3)}\\
\hline
\hline
\textbf{MSSM,}  & $\mathbf{\tan \beta = 30}$  & $\mathbf{\tan \beta = 50}$ \\ 
 $m_{A} = 130$\,GeV             & $m_{h} = 122.7$\,GeV & $m_{h} = 124.4$\,GeV\\
                                & $m_{H} = 134.2$\,GeV & $m_{H} = 133.5$\,GeV\\
\hline
$\sigma \times {\rm BR (A\rightarrow b\bar{b})}$ & 0.07\,fb & 0.2\,fb \\
$\sigma \times {\rm BR (h\rightarrow b\bar{b})}$ & 5.6\,fb  & 13\,fb \\
$\sigma \times {\rm BR (H\rightarrow b\bar{b})}$ & 8.7\,fb  & 23\,fb \\
\hline
\hline
\textbf{MSSM,}  & $\mathbf{\tan \beta = 30}$  & $\mathbf{\tan \beta = 50}$ \\ 
 $m_{A} = 100$\,GeV             & $m_{h} = 98$\,GeV & $m_{h} = 99$\,GeV\\
                                & $m_{H} = 133$\,GeV & $m_{H} = 131$\,GeV\\
\hline
$\sigma \times {\rm BR (A\rightarrow b\bar{b})}$ & 0.4\,fb & 1.1\,fb \\
$\sigma \times {\rm BR (h\rightarrow b\bar{b})}$ & 70\,fb  & 200\,fb \\
$\sigma \times {\rm BR (H\rightarrow b\bar{b})}$ & 8\,fb  & 15\,fb \\
\hline
\end{tabular}
}
\vspace*{-2mm}
\end{table}
At a later stage it might even be possible for TOTEM +CMS to observe exclusive 
production of the Higgs boson. However, the low cross-section requires running at
$\mathcal{L}\sim 10^{33}\,\rm cm^{-2}s^{-1}$, i.e. with scenario 5 whose optics
are such that additional Roman Pots in the cryogenic LHC region at 420\,m from the IP
would be needed for sufficient leading proton acceptance. Still, the
diffractive production rate of a Standard Model Higgs is very low, as is the 
signal-to-background ratio for the dominant decay channel 
$\rm H \rightarrow b\bar{b}$ (see Table~\ref{tab_higgs}, top block).
More favourable is the MSSM case, particularly for large $\tan \beta$ and low 
$m_{A}$ (Table~\ref{tab_higgs}, middle and bottom blocks). Due to the selection
rules~(\ref{eqn_selectionrules}), exclusive production of the CP-odd A is suppressed,
giving the opportunity to separate it from the CP-even h and H, which is difficult
for conventional inclusive production, particularly in the region of 
$m_{A}\approx$\,130\,GeV where all three neutral Higgs bosons have very similar masses.


\end{document}